\documentclass[aps,amsmath,amssymb,prl,twocolumn,footinbib]{revtex4-1}

\usepackage{amsmath}
\usepackage{amssymb}
\usepackage{lineno}
\usepackage{lipsum}
\usepackage{graphicx}
\usepackage{dcolumn}
\usepackage{bm}
\usepackage{color}
\usepackage{xcolor}
\usepackage{mathtools}
\usepackage{appendix}
\usepackage{textcomp}
\usepackage[outdir=./]{epstopdf}

\usepackage[colorlinks, allcolors=blue]{hyperref}

\begin{document}

\title{Information Flow as an Emergent Property of Divergence in Phase-Space}



\author{Praveen Kumar}
\email[]{kumar1@illinois.edu}
\affiliation{Department of Civil and Environmental Engineering, \\ University of Illinois at Urbana-Champaign, Urbana, Illinois 61801, USA}


\date{\today}

\newcommand{\I}[1][mn]{\mathcal{J}_{#1}}
\newcommand{\T}[1][mn]{\mathcal{T}_{#1}}
\newcommand{\D}[1][mn]{\mathcal{D}_{#1}}
\newcommand{\Is}{\vec{V}}
\newcommand{\Ds}{\vec{W}_\tl}
\newcommand{\Fmn}[1][]{\vec{F}_{#1}}
\newcommand{\R}[1]{R^{#1}}
\newcommand{\G}{\mathcal{G}}
\newcommand{\Ss}[1]{S^{#1}}
\newcommand{\U}[2]{U^{#1}_{#2}}
\newcommand{\Bmn}[1]{\vec{X}^{mn}_{#1}}
\newcommand{\Bm}[1]{\vec{X}^{m}_{#1}}
\newcommand{\Bn}[1]{\vec{X}^{n}_{#1}}
\newcommand{\Br}[1]{\vec{X}^{\text{rest}}_{#1}}
\newcommand{\Tar}[1]{X^{\text{tar}}_{{#1}}}
\newcommand\tl{{\tau_\mathbb{C}}}
\newcommand{\verteq}{\rotatebox{90}{$\,=$}}

\newcommand{\Z}{Z}

\newcommand{\p}{\mathit{p}}

\newcommand{\F}{F}
\newcommand{\E}{\mathcal{E}}

\newcommand{\Fv}{\underbar{F}}
\newcommand{\Zv}{\underbar{Z}}

\newcommand{\MVI}{M_{VI}}

\begin{abstract}
Recent developments have created the ability to quantify information flow among  components that interact in a dynamical system, and have led to significant advances in characterizing the dependence between the variables involved. In particular, they have been used to  characterize causal dependency and feedback using observations across diverse fields such as  environment, climate, finance, and human health. What causes information flow among coupled components of a dynamical system? This fundamental question has remained unanswered so far. Here it is established that the information flow is an emergent response  resulting from the divergence of trajectories in phase-space of a dynamical system. This finding shows that the dynamics encapsulated in  the traditional expression of Liouville equation, which neglects this divergence, merely propagates the dependence encoded in the initial conditions. However, when this is not the case, the informational dependence between the components change creating an information flow. This finding has significant implications in a variety of fields, both for the interpretation of observational data for causal inference in natural dynamics, and design of systems with targeted informational dependency.

\end{abstract}


\maketitle

\section{Introduction}

Dynamics in natural systems, such as those associated with the environment, climate, brain etc. exhibit a range of emergent responses arising as a result of interdependencies between interacting components. The dynamical representations of these systems often capture the coupling between components through force balance and/or conservation laws such as those for mass, momentum, and energy. However, the interdependencies also reflect information propagation between system components, as fluctuations in one component drive those in others. We  characterize this exchange as flow of information since the pattern of variability, or uncertainty, in one variable shapes the variability in the coupled variable \cite{GoodwellPNAS2018}. Thus, information flow, quantified as  uncertainty-reducing, or predictive knowledge from one variable to another \cite{ Ruddell2009a}, serves as the currency of exchange between these interacting variables.

Quantifying information flow provides a powerful approach for understanding and characterizing the dependence among components in a variety of physical systems \cite{GoodwellPNAS2018, Sendrowski2018, Franzen2020}. Empirical characterizations of information flow using observed data through measures based on transfer entropy  \cite{Schreiber2000} in a two-way dependence \cite{Gencaga2015, Runge2012, RungeScience2019}, or pairwise dependence in a network of interacting variables \cite{Ruddell2009b}, have become a standard approach for Granger causality based inference  \cite{GoodwellDebate2020} and offer significant possibilities for understanding the behavior of natural systems. More recently, partial information decomposition has offered a more refined way to characterize dependence in a network of interacting variables through a systemic view \cite{Goodwell2017a, Goodwell2017b} or  through their temporal evolution represented using directed acyclic graphs \cite{Runge2012, Runge2015, Jiang2018, Jiang2019, Jiang2020}.

However, a central question still remains unanswered - what causes information flow among coupled variables in a dynamical system? That is, what attributes of a dynamical system give rise to information flow among the set of variables involved? Answering these questions will provide a  foundational perspective for understanding the behavior of natural systems. We address them by identifying the basis of  information flow in dynamical systems. We derive general results for continuous time multivariate autonomous  systems, and specific results associated with multivariate interactions in two- and three-variable systems. 

Our results  below establish the important role played by  the divergence of trajectories in phase-space  \cite{PhaseSpace} in shaping information flow among component variables. These formulations expand upon the Liouville representation of densities  associated with divergence-less flows. They also augment the generalized Liouville representation \cite{Steeb1979, Steeb1980} that was aimed at overcoming these limitations and associated entropy dynamics \cite{Andrey1985, Ramshaw1986, Ezra2004}.  In particular, they  draw out the dependence structure through explicit formulation of the dynamics of multivariate dependence with that of bivariate mutual information as a special case. In commonly used Liouville representation associated with dynamical systems \cite{LiangKleeman1,LiangKleeman2} which neglects the divergence in phase-space, we show that entropic structure encapsulated in the initial conditions is merely advected and not altered through the dynamics. However, when the  divergence of the flow field in the phase-space is non-zero, the entropic dependence changes and drives information flow among the system variables.

Since we use variables in continuous time, entropy is interpreted as differential entropy or may be considered in the context of quantization of the variable involved. However, this limitation is of no practical consequence when mutual information or other multivariate dependence is considered (see chapter 9 in \cite{Cover2006}). As such the results derived here are broadly applicable. 

\section{Probability Density in Phase-Space}

To approach our key question, we first develop the equation governing the dynamics of the multivariate probability distribution of a
 system. This is then used to derive the dynamical equations for the joint and marginal entropies along with the mutual information between the variables. These equations then provide the insights regarding information flow among the system components. We consider a system consisting of $N$ variables $\Zv(t) \equiv \left[\Z_1(t), \Z_2(t),  \dots, \Z_N(t)\right]$, with $\Z_i(t)$ defined on the support $\Omega_i$.   Consider its dynamics given as:
\begin{equation}
 	\dot{\Zv}(t) \equiv \frac{d\Zv(t)}{dt}= \Fv(\Zv(t))   \label{eq:Dyn}
\end{equation}
where $\Zv(t) \in \Omega$ with $\Omega = \Omega_1 \times \Omega_2 \times \dots \times \Omega_N$. $\Fv(\Zv)  \equiv \left[ \F_1(\Zv), \F_2(\Zv),  \dots, \F_N(\Zv)\right]$ where the function $\F_i(\Zv)$ captures the dynamics of the individual components as a function of all variables, that is, $d\Z_i(t)/dt = F_i(\Zv(t))$. Let us consider the representation in the phase-space, that is, the space of coordinates introduced by the components  $\Z_i$.   We explore the probability of finding a trajectory in any differential volume $d\Omega$ at time $t$. A practical approach to obtain this probability is by considering a large number of trajectories, starting with random initial conditions. The fraction of these trajectories that pass through $d\Omega$ at time $t$ provide an estimate of the probability density function ({\it pdf}) $\p(\Zv, t)$ with $\int_{\Omega} \p(\Zv, t) \, d\Zv = 1$.  Equivalently, we may consider $\p(\Zv,t)$ as a density field in phase-space through which the trajectories traverse. We assume trajectories are distinct and they are neither created nor destroyed.  We also assume that $\p(\Zv, t)$ has a compact support over $\Omega$ or decays exponentially fast.

\begin{figure*}[tb]
 \centering
 \includegraphics[width=\textwidth]{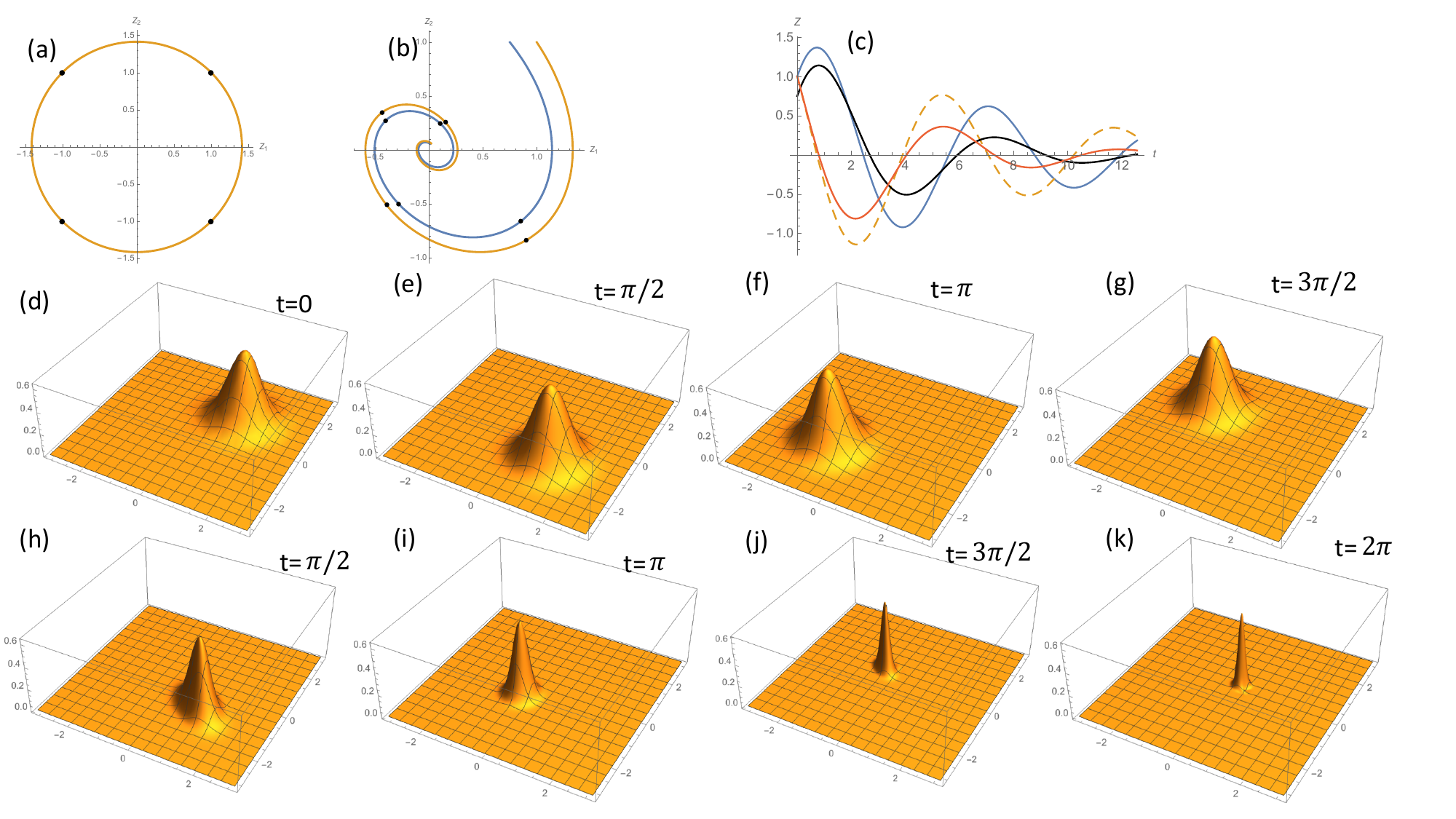}
 \caption{Illustration of the role of $\nabla \cdot \Fv$ in  the  dynamics of {\it pdf} for a damped harmonic oscillator (equation (\ref{eq:dampedHS})) in comparison to an undamped case. (a) Phase-space plot of an undamped, and (b) damped harmonic oscillator (with two nearby trajectories). The dots indicate position at times in multiples of $\pi/2$ after the initial time. Subplot (c) shows the time series of position ($\Z_1$) [blue, grey] and velocity ($\Z_2$) [red, orange] corresponding to the two trajectories in (b). Subplot (d) shows the initial condition for $p$ (product of two independent univariate Gaussian distributions with equal variance (0.25)). Subplots (e), (f) and (g) show the evolution of $p$, for the undamped case with $b=0$, therefore corresponding  to equation (\ref{eq:Lville}),  at times $\pi/2$, $\pi$, and $3\pi/4$ from initial time. At $t = 2\pi$ the systems returns to that in (d). Starting with the initial condition in (d), subplots (h), (i), (j), and (k) show evolution of $p$ for $b/m=1/2$, thereby corresponding to equation (\ref{eq:pdiv2}) with $\nabla \cdot \Fv \neq 0$, at times $\pi/2$, $\pi$, $3\pi/4$ and $2\pi$ respectively.   (color online) }
 \label{fig:phase-pdf}
\end{figure*}

By considering the  total derivative of a trajectory $\frac{d \p}{dt} =  \frac{\partial \p}{\partial t} + \sum_i \frac{\partial \p}{\partial \Z_i} \frac{d\Z_i}{dt} $  we get (see also Appendix A):
\begin{equation}
	\frac{d \p}{dt}  =   \frac{\partial \p}{\partial t} + \nabla p \cdot \dot{\Zv}(t) \equiv \frac{\partial \p}{\partial t} + \nabla \p \cdot \Fv(\Zv).  \label{eq:dpdt}
\end{equation}
Since all trajectories remain confined within the support $\Omega$ by definition, the total probability over $\Omega$ remains unity at any time resulting in $\frac{d}{dt} \int_{\Omega} p(\Zv,t)\, dt = 0$, and therefore we set $\frac{d \p}{dt}=0$ to arrive at 
\begin{equation}
	\frac{\partial \p}{\partial t} + \nabla \p \cdot \Fv = 0,   \label{eq:pdiv1}
\end{equation} 
where the arguments for $\p$ and $\Fv$ have been dropped for brevity but will be expanded when there is a possibility of ambiguity, a practice we will follow throughout. Note that the second term in equation  (\ref{eq:pdiv1})  captures the gradient of the probability density projected along the flow vector in phase space. 
Using the identity $\nabla \cdot (\p\Fv) = \nabla \p \cdot \Fv +\p \nabla \cdot \Fv$, we  equivalently obtain
\begin{equation}
	\frac{\partial \p}{\partial t} + \nabla \cdot (\p\Fv) -  p \nabla \cdot \Fv = 0  \label{eq:pdiv2}
\end{equation} 
which further illustrates the role if the divergence of the flow vector, $\nabla \cdot \Fv$.  
That is, the {\it pdf} changes as a result of both the way in which the trajectories occupy the phase-space at any time, and the way in which the flow field is structured in phase-space.
We note that when $ \nabla \cdot \Fv = 0$, we obtain the standard form of the Liouville equation:
\begin{equation}
	\frac{\partial \p}{\partial t} + \nabla \cdot (\p\Fv) = 0,     \label{eq:Lville}
\end{equation}
which expresses that the volume in phase-space is preserved in the absence of divergence, and we have a conservative system. 

To examine the important role of phase-space divergence in the dynamics of $\p(\Zv, t)$, let us 
consider a prototypical example of damped harmonic oscillator given in the standard form as $m \ddot{x} + b\dot{x} + k x=0$.  Although this example is elementary, it serves to illustrate the role of the divergence in phase space. Using $\Z_1$ and $\Z_2$ for position and velocity, we get  the phase-space dynamics given as 
\begin{eqnarray}\label{eq:dampedHS}
     \dot{\Z_1} & = & \Z_2(t)  \\ \nonumber 
  \dot{\Z_2} & = & -b/m \,\Z_2(t)-k/m \,\Z_1(t),    
\end{eqnarray}
 resulting in $\nabla \cdot \Fv = -b/m < 0$ for $b,m >0$. For $b=0$ the equation corresponds to the simple harmonic oscillator with $\nabla \cdot{\Fv} = 0$, a prototypical example of a conservative system, but  otherwise it corresponds to a dissipative dynamical system. In this particular situation  the trajectories converge closer to each other with time (Fig. \ref{fig:phase-pdf}a,b,c). To illustrate the role of $\nabla \cdot{\Fv}$, we further show the evolution of the {\it pdf} for two situations corresponding to $b=0$ (Fig. \ref{fig:phase-pdf}e-g) and $b/m=1/2$ (Fig. \ref{fig:phase-pdf}h-k) starting with the same initial {\it pdf} (Fig. \ref{fig:phase-pdf}d). For the dissipative case, as the trajectories close in together (as illustrated in Fig. \ref{fig:phase-pdf}b), the structure of the {\it pdf} is modified. This is in contrast to the conservative case where the {\it pdf}  merely gets advected in phase-space. As a result, in the case of a conservative system, information is conserved over time \cite{henriksson2019gibbsliouville}, that is, the dynamics doesn't create or destroy any information that is not already contained in the initial condition. However, for the dissipative system, the information content changes with time because the entropic behavior of the {\it pdf} changes.

The classic Lorenz equation for deterministic chaos given as 
\begin{eqnarray}
 	 \dot\Z_1 & = & \sigma(\Z_2(t) - \Z_1(t)) \\ \nonumber
	 \dot{\Z_2} & = & \Z_1(t) (\rho-\Z_3(t))-\Z_2(t) \\ \nonumber
	 \dot{\Z_3} & = & \Z_1(t) \Z_2(t) - \beta \Z_3(t)     \label{eq:Lorenz}
\end{eqnarray}
results in $\nabla \cdot{\Fv} = -(\sigma + \beta + 1) < 0 $ for usual parameters $\sigma, \beta, \rho > 0$, and serves as another important example of a dissipative system. The phase-space changes its structure, and volumes in phase-space are not conserved with the evolution of the system, thereby making the use of the standard Liouville equation (\ref{eq:Lville})  inadmissible for its exploration or other such systems.

To the best of author's knowledge, the general form in equation  (\ref{eq:pdiv2}) (or equation (\ref{eq:pdiv1})) has not been previously considered in characterization of information flow in dynamical system. Indeed the  work presented by \cite{LiangKleeman1, LiangKleeman2} is based on Liouville equation (\ref{eq:Lville}) which is formulated based on the underlying assumption of $\nabla \cdot \Fv=0$, thereby excluding the impact of the divergence of the phase-space on the probability density $\p(\Zv,t)$ (as illustrated in Fig. \ref{fig:phase-pdf}). 

This brings us to the key tenet of this work. From equation (\ref{eq:pdiv2}) we note that the change in $\p(\Zv, t)$  is a balance between the divergence of trajectories resulting from the divergence of the flow field in the phase-space. One way to interpret the initial  {\it pdf}, $p(\Zv, 0)$, is to think of it as representing the probability of a selection (or ensemble) of trajectories whose dynamics we wish to explore. As the system evolves,  the phase-space volume occupied by the trajectories is preserved when $\nabla \cdot \Fv = 0$ and as a result the {\it pdf} is not entropically altered, merely advected in the phase-space. However, when $\nabla \cdot \Fv \neq 0$, the trajectories either diverge or are squeezed together. This is accomplished through the modification of the relationship that the components $\Z_i$'s have with each other within a trajectory as dictated by the structure embodied in the relationship $\Fv(\Zv)$. So while the change in density is associated with the squeezing or expansion of nearby trajectories, this  is a result of the interaction between  the different components comprising the dimensions of the phase-space. Therefore the changing {\it pdf} of the ensemble is a reflection of the changing relationship between the variables in the individual trajectories. In other words, the dynamical relation $\Fv$ induces an informational dependence between the system components $\Z_i$.  This is akin to vehicles squeezing from a closed lane in a multilane highway, and the vehicles in the open lanes slowing down to accommodate the changing pattern of traffic flow drawing upon the information of changing traffic pattern. We can therefore use the dual view for the {\it pdf}, one associated with the ensemble and the other with the changing relation between components of the dynamics. So we interpret the change in the {\it pdf} as a reflection of the changing relation between the components $\Z_i$ in the dynamics. That is, the dynamics of the {\it pdf}, and the informational attributes it encapsulates, is not merely a statistical characterization of the trajectories but a physical attribute of the system behavior itself. We can therefore use this {\it pdf} to characterize the dynamics of entropy and multivariate mutual information among the components $\Z_i$.

\section{Dynamics of Entropy}

We can now use equation (\ref{eq:pdiv2}) to determine the evolution of entropy and explore its dependence on $\nabla \cdot \Fv$ . The dynamics of the system entropy, $H_{\Zv}(t)$, associated with the joint distribution $\p(\Zv, t)$ can be derived as (see Appendix B): 
 \begin{equation}
 	\frac{d H_{\Zv}}{d t}  -  \int_{\Omega}  (\p \log \frac{1}{\p})  \nabla \cdot \Fv \, d\Zv   = 0.   \label{eq:Hdiv}
\end{equation}
Alternatively this may be written as 
 \begin{equation}
 	\frac{d H_{\Zv}}{d t}  -  \E  \left[ \psi(\Zv,t) \nabla \cdot \Fv \right]    = 0    \label{eq:Hdiv1}
\end{equation}
where $\E$ is the expectation operator and $\psi(\Zv,t) = \log (1/\p(\Zv,t))$ is the pointwise information in the phase-space, such that $H_{\Zv}(t) = \E [\psi(\Zv,t)]$. This equation immediately draws out the crucial role of  $\nabla \cdot \Fv$ in the evolution of the system entropy.

For the situation when $\nabla \cdot \Fv$ is independent of $\Zv$, i.e. it invariant in the phase-space, for example as in the case of the damped harmonic oscillator or the Lorenz system, we have 
\begin{equation}
    \E  \left[ \psi(\Zv,t) \nabla \cdot \Fv \right] = ( \nabla \cdot \Fv) H_{\Zv}
\end{equation}
and equation (\ref{eq:Hdiv1}) gives us the dynamics of the system entropy as, 
\begin{equation}
	\frac{d H_{\Zv}}{d t}  -  \left(\nabla \cdot \Fv \right) H_{\Zv}  =  0.    \label{eq:Hdiv2}
\end{equation}
This equation admits a direct solution 
\begin{equation}
 	H_{\Zv} (t) = H_{\Zv}( t_0) \exp{\left\{(\nabla\cdot \Fv)\Delta t \right\}}
\end{equation}
where $\Delta t = t-t_0$ with $H_{\Zv}( t_0)$ being the entropy at the initial time $t_0$. We see that the system is entropically altered by phase-space divergence during its evolution.

We note that for a conservative system governed by Liouville equation (\ref{eq:Lville}) we get	$ H_{\Zv}(t) = H_{\Zv}(t_0)$ 
reflecting that entropy is temporally invariant and no information is generated by the dynamics, consistent with known understanding. So while for a simple harmonic oscillator the entropy is constant, for a damped harmonic oscillator it decays as 
$H(t) = H(t_0) \exp{\left\{-(b/m)\Delta t \right\}}$, and  for the Lorenz system it varies as
\begin{equation}
H_{\Zv}( t) = H_{\Zv} (t_0) \exp{\left\{-(\sigma + \beta + 1)\Delta t \right\}}. \label{eq:LorenzH}
\end{equation}

There are situations when $\nabla \cdot \Fv$ is not invariant in the phase-space, In such cases 
 $\psi(\Zv,t)$ plays an important role. An example is provided by the R$\ddot{\rm o}$ssler system given as
\begin{eqnarray}
	\dot{\Z_1} &= & -\Z_2(t) -\Z_3(t) 	\\ \nonumber
	\dot{\Z_2} & = & \Z_1(t) + a\Z_2(t) 	\\ \nonumber
	\dot{\Z_3} & = & b+\Z_3(t)(\Z_1(t)-c)
\end{eqnarray}
where $a,b$ and $c$ are parameters. It is easily seen that $\nabla \cdot \Fv = a - c + \Z_1$ and $\E  \left[ \psi(\Zv,t) \nabla \cdot \Fv \right]  = (a-c)H_{\Zv}+ \int_\Omega (p \log(1/p) \Z_1\, d\Zv$, which appears more complex than that for the Lorenz system.

Equation (\ref{eq:Hdiv1}) (or equation (\ref{eq:Hdiv})) is the key result that characterizes the evolution of the system entropy and shows that phase-space divergence is the primary determinant of this dynamics.  
We can now use  equation (\ref{eq:Hdiv1})  for the joint entropy  to characterize multivariate interaction between the system components.

\section{Dynamics of Multivariate Interaction}

To understand how mutual information and higher dimensional multivariate interactions evolve, we invoke
 the chain rule for entropy, i.e., $H_{\Z_1, \cdots, \Z_N} (t) = H_{\sum_{i=1}^N\{ \Z_i | Z_{i-1}, \cdots, \Z_1\}} (t)$,  and by substituting in equation (\ref{eq:Hdiv1}) we get
\begin{equation}
	\sum_{i=1}^N \frac{\partial}{\partial t} H_{ \Z_i | Z_{i-1}, \cdots, \Z_1} -  \E  \left[ \psi(\Zv,t) \nabla \cdot \Fv \right]    = 0.  \label{eq:ChainRule}
\end{equation}
For a 2-variable case, using equation (\ref{eq:ChainRule}) we can show that the mutual information evolves as a function of the marginal entropies as (see Appendix C) :
\begin{equation}
	 \frac{\partial I_{\Z_1;\Z_2}}{\partial t}   =   \frac{\partial}{\partial t} \left( H_{\Z_1}   + H_{\Z_2}  \right)   -  \E  \left[ \psi(\Zv,t) \nabla \cdot \Fv \right] .   \label{eq:MI2var1}
\end{equation}
This equation again encapsulates the  contribution of phase-space divergence in the evolution of dependence in a bivariate system. 

For the special case when $\nabla \cdot \Fv $ is independent of $\Zv$, i.e. equation (\ref{eq:Hdiv2}) holds,  from equation (\ref{eq:MI2var1}) we get
\begin{equation}
	 \sum_{i=1}^2  \left( \frac{\partial H_{\Z_i}}{\partial t} - (\nabla \cdot F)H_{\Z_i} \right) -    
	 \left(\frac{\partial I_{\Z_1;\Z_2}}{\partial t} - (\nabla \cdot F) I_{\Z_1;\Z_2}\right) =  0.
\end{equation}
This equation links the dynamics of the marginal entropies  of $\Z_1$ and $\Z_2$ to the dynamics of their mutual information.  For conservative systems, using $\nabla \cdot \Fv = 0$, we easily get 
 \begin{equation}
 	\frac{\partial}{\partial t} \left( H_{\Z_1} + H_{\Z_2} - I_{\Z_1;\Z_2} \right) = 0
 \end{equation}
 consistent with $H_{\Z_1,\Z_2} (t)$ remaining invariant with $t$ although the balance between the marginal entropies and mutual information can change in time. 
 
We now consider dependence between three variables, for which we have  $H_{\Z_1,\Z_2, \Z_3} (t) = H_{\Z_1} (t) + H_{\Z_2|Z_1}(t) + H_{\Z_3|\Z_2,\Z_1}(t) $ from the chain rule. This gives us, from equation (\ref{eq:ChainRule}),
\begin{equation}
 	\frac{\partial}{\partial t} \left( H_{\Z_1} + H_{\Z_2|Z_1} + H_{\Z_3|\Z_2,\Z_1} \right) - \E  \left[ \psi(\Zv,t) \nabla \cdot \Fv \right]    = 0.
\end{equation}
By noting the following identities $H_{\Z_i | \Z_j} = H_{\Z_i} - I_{\Z_i;\Z_j}$ and  $H_{\Z_3|\Z_2,\Z_1} = H_{\Z_3|\Z_2} - I_{\Z_3;\Z_1 | \Z_2}$  the above equation can be written as

\begin{eqnarray}
 	\frac{\partial}{\partial t} \left( H_{\Z_1} + H_{\Z_2} + H_{\Z_3} \right) & -&  \\ \nonumber
	     					 \frac{\partial}{\partial t} \left( I_{\Z_1;\Z_2} + I_{\Z_2;\Z_3}  +  I_{\Z_3;\Z_1 | \Z_2} \right) & - &  \E  \left[ \psi(\Zv,t) \nabla \cdot \Fv \right]     = 0.
\end{eqnarray}

Noting further that interaction information is given as $ I_{\Z_1;\Z_2;\Z_3} =  I_{\Z_1;\Z_2 } - I_{\Z_1;\Z_2|\Z_3 } = I_{\Z_2; \Z_3}  - I_{\Z_2;\Z_3|\Z_1 }$ we get two equivalent forms involving multivariate information, $\MVI(\Zv)$, that captures the dependence among the variables:
\begin{eqnarray}
 		 \MVI(\Zv) & = & I_{\Z_1;\Z_2} + I_{\Z_2;\Z_3}  +  I_{\Z_3;\Z_1 } - I_{\Z_1;\Z_2;\Z_3}    \\ \nonumber
		          & = &  I_{\Z_1;\Z_2|\Z_3} + I_{\Z_2;\Z_3|\Z_1}  +  I_{\Z_3;\Z_1 | \Z_2} +2 I_{\Z_1;\Z_2;\Z_3}      \label{eq:MVI1}
\end{eqnarray}
and
\begin{equation}
 		 \frac{\partial \MVI}{\partial t}   
	 = \left( \sum_{i=1}^3 \frac{\partial}{\partial t} H_{\Z_i} \right)-     \E  \left[ \psi(\Zv,t) \nabla \cdot \Fv \, \right]    \label{eq:MVI2}
\end{equation}
where, akin to equation (\ref{eq:MI2var1}), the LHS characterizes the dynamics of the interaction between the  variables. We again note that $\nabla \cdot \Fv$ asserts an important role in the evolution of the multivariate interaction information. For the special case when $\nabla \cdot \Fv$ is independent of $\Zv$, equation (\ref{eq:MVI2})  reduces to
\begin{equation} \label{eq:MVI3}
 		  \frac{\partial \MVI }{\partial t}   
		 -\left(\nabla \cdot \Fv \right) \left( \MVI \right)  
	 = \sum_{i=1}^3 \left( \frac{\partial H_{\Z_i} }{\partial t}  -       \left( \nabla \cdot \Fv \right) H_{\Z_i} \right).   
\end{equation}

Since the multivariate interaction information, $\MVI$, is a function of the marginal entropies, it is possible to  obtain explicit equations for the evolution of the marginal entropy $H_{\Z_i}(t)$ as (see Appendix D):
\begin{equation}
	 \frac{\partial H_{\Z_i}}{\partial t}   =   \int_{\Omega} \log \p_i \frac{\partial(p\F_i)}{\partial \Z_i} \, d\Zv_i   
	                                                                                                                      -   \E \left[(1-\psi(\Z_i,t))\nabla \cdot \Fv \right],  \label{eq:MarginalH1}
\end{equation} 
where $\psi(\Z_i,t) = \log \frac{1}{p_i}$  with the property that $\E \left[ \psi(\Z_i, t) \right] = \int_{\Omega} p \log \frac{1}{p_i} d\Zv = \int_{\Omega_i} p_i \log \frac{1}{p_i} d\Z_i = H_{\Z_i}$.
This formulation for the marginal entropy together with that for the system (equation  (\ref{eq:Hdiv}) or (\ref{eq:Hdiv1})) allows us to completely characterize the dynamics of the multivariate dependence.
 
 For the two variable case, inserting equation (\ref{eq:MarginalH1})  into equation (\ref{eq:MI2var1}), we get 
\begin{equation}
	 \frac{\partial I_{\Z_1;\Z_2}}{\partial t}    =      \sum_{i=1}^{2} \left[ \int_{\Omega} \log \p_i  \frac{\partial (p\F_i)}{\partial \Z_i} \, d\Z_i  \right]      
	                                                           -   		 (2+    I_{\Z_1;\Z_2} )  \, \nabla \cdot \Fv 		\label{eq:MutualInfo2var}
\end{equation}
Similarly for a three variable case, using equation (\ref{eq:MarginalH1}) in equation (\ref{eq:MVI2}), we get 
\begin{equation}
	 \frac{\partial \MVI}{\partial t}    =      \sum_{i=1}^{3} \left[ \int_{\Omega} \log \p_i  \frac{\partial (p\F_i)}{\partial \Z_i} \, d\Z_i  \right]      
	                                                           -   		 (3+    \MVI)  \, 	\nabla \cdot \Fv	\label{eq:MutualInfo3var}
\end{equation}
The above two equations serve to illustrate how multi-variate interactions between variables are shaped by both the dynamical relations captured in $F(\Zv)$ as well its divergence in phase-space. In the form of equations (\ref{eq:MutualInfo2var} and \ref{eq:MutualInfo3var}), the dynamics of the interaction can be directly computed without the need to compute the marginal and joint entropies.


\section{Conclusion}

Our key results are encapsulated in the dynamics of the {\it pdf} (equations \ref{eq:pdiv1} or \ref{eq:pdiv2}), joint entropy (equations \ref{eq:Hdiv} or \ref{eq:Hdiv1}),  marginal entropy (equation \ref{eq:MarginalH1}), and multivariate interaction information (equation \ref{eq:ChainRule}). Based on these, specific results for the dynamics of the bivariate mutual information (equation (\ref{eq:MutualInfo2var})) and trivariate interaction information (equation (\ref{eq:MutualInfo3var})) are established. Based on the insights gained from these results  we  conclude that the divergence of the flow field in the phase-space  alters the entropic structure of the {\it pdf}, that is, it creates temporal information change, during the evolution of a dynamical system. As a result it induces information flow among the component variables involved. If this divergence is zero, as is the case with the traditional implementation associated with Liouville equation, we simply propagate the dependence embodied in the initial conditions. These results provide a foundational basis for thinking about the evolving dynamics of information flow and have potential applications in many fields. In particular, in the study of natural phenomena, such as those associated with environmental and climatic systems, these results provide the potential to explore the basis of evolution of dependence among interacting variables. 

While the results include expression only for temporally synchronous dependence through information flow, we can esily envision that time-lagged dependence between the system components, such as those sought through transfer entropy \cite{Schreiber2000}, also change as a result. These will be explored in a future studies. 


\appendix

\section{Appendix A} 
\label{sec:appendixA}

\begin{figure}[tb]
 \centering
 \includegraphics[trim=0mm 70mm 0mm 00mm, clip, width=\textwidth]{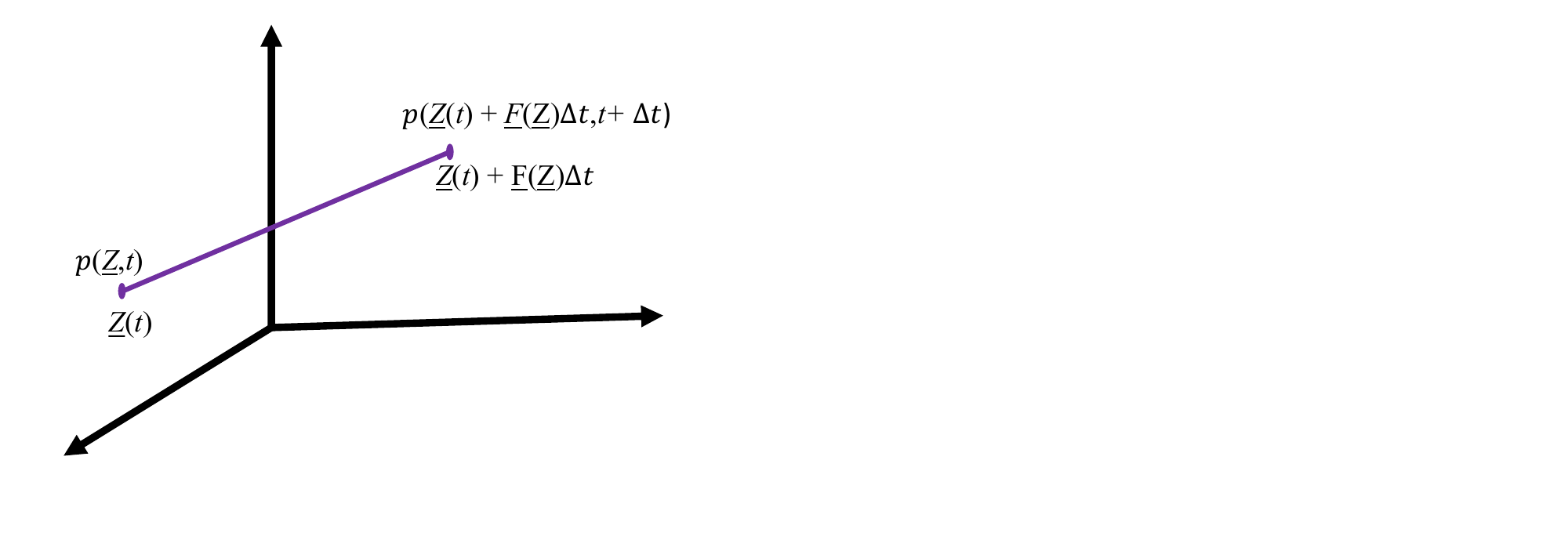}
 \caption{Two points in phase space separated $\Delta t$ time apart at $\Zv(t)$ and $\Zv(t) + \Fv(\Zv)\Delta t$ have  associated {\it pdf}  $p(\Zv,t)$ and $p(\Zv + \Fv \Delta t, t+\Delta t)$. }
 \label{fig:VectorRep}
\end{figure}

Here we provide an alternate derivation of equation (\ref{eq:dpdt}) by considering the {\it pdf} $p(\Zv,t)$ of a trajectory in phase space (see Figure \ref{fig:VectorRep}). Consider the Taylor series expansion  about $(\Zv,t)$:
\begin{eqnarray}
p(\Zv + \Fv \Delta t, t+\Delta t) & = & p(\Zv,t) + \frac{\partial p(\Zv,t)}{\partial t} \Delta t \\ \nonumber
                                                 & + & \nabla p(\Zv,t) \cdot \Delta \Zv + \textrm{higher order terms}.
\end{eqnarray}
Therefore, by noting $\lim_{\Delta t \to 0} \Delta Z/\Delta t = \Fv(\Zv)$ and neglecting higher order terms, the Lagrangian derivative is given as
\begin{eqnarray}
\frac{d p}{d t}  & = &  \lim_{\Delta t \to 0} \frac{p(\Zv + \Fv \Delta t, t+\Delta t) - p(\Zv,t)}{\Delta t} \\ \nonumber
                        & = & \frac{\partial p(\Zv,t)}{\partial t} + \nabla p(\Zv,t) \cdot  \Fv(\Zv)
\end{eqnarray}
where the terms on the RHS comprise the Eulerian derivative. 

\section{Appendix B} 
\label{sec:appendixB}

Here we show the derivation of equation (\ref{eq:Hdiv}). We multiply  equation (\ref{eq:pdiv1}) with $1+\log\p$ to get 
\begin{equation}
	(1+\log\p) \frac{\partial \p}{\partial t} + (1+\log\p) \nabla \p \cdot \Fv = 0,   
\end{equation} 
Noting that $(1+\log\p) \frac{\partial \p}{\partial t} = \frac{\partial\p\log\p}{\partial t}$, and further expanding $\nabla \p \cdot \Fv$ and $\log\p \nabla \p \cdot \Fv$ and adding individual term we get $(1+\log\p) \nabla \p \cdot \Fv = \sum_i \F_i (1+\log\p)\frac{\partial \p}{\partial \Z_i} =  \sum_i \F_i \frac{\partial (\p \log \p)}{\partial \Z_i}$ giving us 
\begin{equation}
     \frac{\partial (\p \log \p)}{\partial t} + \nabla (\p \log \p) \cdot \Fv = 0
 \end{equation}
which can be written as 
\begin{equation}
     \frac{\partial (\p \log \p)}{\partial t} + \nabla \cdot (\p \log \p \, \Fv) - (\p \log \p) \nabla \cdot \Fv = 0.
 \end{equation}
 Multiplying by $-d\Zv$ and integrating over $\Omega$ we get
 \begin{eqnarray}
 	\int_{\Omega} \frac{\partial} {\partial t} (\p \log \frac{1}{\p}) d\Zv & + &  \int_{\Omega}    \nabla \cdot (\p \log \frac{1}{\p} \, \Fv) d\Zv \\ \nonumber
	                                                                                                            & - & \int_{\Omega}  (\p \log \frac{1}{\p})  \nabla \cdot \Fv \, d\Zv   = 0 . 
\end{eqnarray}
The first term is $\frac{\partial H_{\Zv}}{\partial t} =  \frac{\partial} {\partial t} \int_{\Omega} (\p \log \frac{1}{\p}) d\Zv$ where $H_{\Zv}(t)$ is the Shannon entropy associated with the joint distribution over the phase-space $\Zv$ at time $t$. To evaluate the second term, we invoke the divergence theorem to get 
$\int_{\Omega}    \nabla \cdot (\p \log\frac{1}{\p} \, \Fv) d\Zv = \int_{\delta\Omega} ( \p \log\frac{1}{\p}) \, \Fv \cdot \vec{n} \, ds$ where ${\delta\Omega}$ represents the surface for the domain $\Omega$, $ds$ is a differential element on this surface and $\vec{n}$ is the normal to this surface. Since the flux of probability through this surface is of measure zero, this term is zero and we get equation (\ref{eq:Hdiv}).


\section{Appendix C} 
\label{sec:appendixC}
Here we show the derivation of equation (\ref{eq:MI2var1}). Equation (\ref{eq:ChainRule}) can be written in terms of marginal entropies and multivariate interaction. Consider a 
 two variable case, where $H_{\Z_1,\Z_2} (t) = H_{\Z_1} (t) + H_{\Z_2|\Z_1}(t) $. Noting that $H_{\Z_2|\Z_1}(t) = H_{\Z_2}(t) - I_{\Z_1;\Z_2} (t)$ where $I_{\Z_1;\Z_2}(t) $ is the mutual information between $\Z_1(t)$ and $\Z_2(t)$, we get
\begin{equation}
	\frac{\partial}{\partial t} \left( H_{\Z_1}   + H_{\Z_2}  -  I_{\Z_1;\Z_2} \right)   -  \E  \left[ \psi(\Zv,t) \nabla \cdot \Fv \right]    = 0.
\end{equation}
Alternatively, this can be written to show that the mutual information evolves as a function of the marginal entropies as:
\begin{equation}
	 \frac{\partial I_{\Z_1;\Z_2}}{\partial t}   =   \frac{\partial}{\partial t} \left( H_{\Z_1}   + H_{\Z_2}  \right)   -  \E  \left[ \psi(\Zv,t) \nabla \cdot \Fv \right] .   \label{eq:MI2var2}
\end{equation}
This gives equation (\ref{eq:MI2var1}).


\section{Appendix D} 
\label{sec:appendixD}

Here we show the derivation of equation (\ref{eq:MarginalH1}). 
We note that the marginal entropy $H_{\Z_i}(t) = \int_{\Omega_i} \p_i \log(1/\p_i) \, d\Z_i$ where $\p_i(t)$ is the marginal {\rm pdf} is obtained as
\begin{equation}
	\p_i(t) = \int_{\Omega \setminus \Omega_i} \p(\Zv,t) \, d\Zv  \label{eq:pi}
\end{equation}
and where $\setminus$ is the exclusion operator and $d\Zv$ in this case is understood to exclude $d\Z_i$ from the context of integral over $\Omega \setminus \Omega_i$. We first integrate equation (\ref{eq:pdiv2}) over the subspace $\Omega \setminus \Omega_i$  as: 
\begin{eqnarray}
	 \frac{\partial}{\partial t} \int_{\Omega \setminus \Omega_i}  \p(\Zv,t) \, d\Zv & +  & \int_{\Omega \setminus \Omega_i} \nabla \cdot (\p\Fv) \, d\Zv    \\ \nonumber
	                                                                                                                     & -  &  \int_{\Omega \setminus \Omega_i} p \nabla \cdot \Fv \, d\Zv = 0.  
\end{eqnarray} 
For the second term we expand $\nabla \cdot (\p\Fv)$ and integrate each term $j$ first as $\int_{\Omega_j} \frac{\partial(p\F_j)}{\partial \Z_j} \, d\Z_j$. This evaluates to $0$ since $p$ has a compact support.
As a result $ \int_{\Omega \setminus \Omega_i} \nabla \cdot (\p\Fv) \, d\Zv = \int_{\Omega \setminus \Omega_i} \frac{\partial(p\F_i)}{\partial \Z_i} \, d\Z_i $ as all other terms are $0$. Further using equation (\ref{eq:pi}), the above simplifies to 
\begin{equation}
	 \frac{\partial p_i}{\partial t}   +   \int_{\Omega \setminus \Omega_i} \frac{\partial(p\F_i)}{\partial \Z_i} \, d\Zv_i   
	                                                                                                                      -    \int_{\Omega \setminus \Omega_i} p \nabla \cdot \Fv \, d\Zv = 0.  
\end{equation} 
Multiplying the above by $-(1+\log \p_i)$ and integrating with respect to $\Z_i$ and noting again that $\int_{\Omega_i} \frac{\partial(p\F_i)}{\partial \Z_i} \, d\Z_i = 0$, we get
\begin{equation}
	 \frac{\partial H_{\Z_i}}{\partial t}   =   \int_{\Omega} \log \p_i \frac{\partial(p\F_i)}{\partial \Z_i} \, d\Z_i   
	                                                                                                                      -    \int_{\Omega} (1 - \log \frac{1}{p_i}) p \nabla \cdot \Fv \, d\Zv .  
\end{equation} 
This reduces to equation (\ref{eq:MarginalH1}). 



\vskip 0.25in

\begin{acknowledgments}
    Funding support from the following ARPA-E grant DE-AR0001225, and NSF grants  EAR 1331906, EAR 2012850,  and OAC 1835834 are acknowledged. Special thanks to Peishi Jiang and  Allison Goodwell for providing excellent insights with the derivations and interpretation, and to Francina Dominguez and Hoshin Gupta for broader discussions. 
\end{acknowledgments}

\bibliography{mybibfile}

\end{document}